# Temperature Coefficients of the Raman Peaks for the Single-Layer and Bi-Layer Graphene

I. Calizo and A.A. Balandin[*]

Nano-Device Laboratory, Department of Electrical Engineering

University of California – Riverside, Riverside, California 92521 U.S.A.

F. Miao, W. Bao and C.N. Lau

Department of Physics and Astronomy

University of California – Riverside, Riverside, California 92521 U.S.A.

We carried out micro-Raman spectroscopy of graphene layers over the temperature range from ~80 K to ~370 K. The number of layers was independently confirmed by the quantum Hall measurements and atomic force microscopy. The measured values of the temperature coefficients for the $G$ and $2D$-band frequencies of the single-layer graphene are $-(1.6\pm0.2)\times10^{-2}$cm$^{-1}$/K and $-(3.4\pm0.4)\times10^{-2}$cm$^{-1}$/K, respectively. The $G$ peak temperature coefficient of the bi-layer graphene and bulk graphite are $-(1.5\pm0.06)\times10^{-2}$cm$^{-1}$/K and $-(1.1\pm0.04)\times10^{-2}$cm$^{-1}$/K, respectively.

---

[*] Corresponding author; electronic address (A.A. Balandin): balandin@ee.ucr.edu

Since its recent micromechanical isolation and measurements by Novoselov *et al,*[1-2] graphene has attracted major attention from the physics and device research communities.[1-12] In addition to the wealth of the two-dimensional (2D) electron gas physics it reveals, graphene has shown promise as a material for the electronic applications beyond the conventional complementary metal-oxide semiconductor (CMOS) technology.[9] It was demonstrated that Raman spectroscopy can serve as a convenient technique for identifying graphene.[13-14] Ferrari *et al*[13] studied the evolution of the *2D* band Raman signatures with the addition of each extra layer of graphene and explained it with the double-resonance model. Gupta *et al*[14] have shown that the *G* peak position $\omega_G$ is sensitive to the number of layers *n*, i.e. $\omega_G \sim 1/n$.

The Raman spectroscopy studies of graphene[13-16] reported to date were limited to the room temperature. In order to expand the use of Raman spectroscopy as a nanometrology tool one has to investigate the change in the Raman signatures of graphene with temperature. This is important for the graphene-based devices, since the application of electric bias and gate voltages results in the device self-heating. Although the thermal conductivity of graphene is expected to be high, there are always thermal resistances associated with the contacts and interfaces between different materials,[17-18] which may lead to the local temperature increase due to excitation or bias affecting the Raman spectrum.[19] The nano-ribbons made of graphene for inducing the band gap may also deteriorate the thermal conductivity via the phonon – boundary scattering and phonon confinement.[20-21] These considerations provided the motivation for the temperature study of the graphene Raman signatures.

Here we report the variable temperature spectroscopic Raman microscopy of the single-layer graphene (SLG) and bi-layer graphene (BLG) deposited on silicon substrates for fabrication of the graphene-based devices. SLG and BLG were obtained by micromechanical cleavage of bulk graphite using the process outline in Refs.[1-2] Before performing the micro-Raman spectroscopy we carried out transport studies to confirm the quality of graphene and verify the number of layers. To characterize the graphene layers and devices, we attached the electrodes to a number of SLG and BLG using the standard nanofabrication techniques (the drain-source separation is 1-5 μm). The electrical measurements were performed at low temperature in a sorption pumped [3]He refrigerator.

A typical piece of graphene selected for fabrication is shown in Fig. 1 (a). Figure 1 (b) shows a characteristic linear – response conductance $g_m$ of the graphene device as a function of the gate voltage $V_g$, which is used to tune the density of the charge carriers $n_s$ in graphene. Assuming a parallel plate capacitance between graphene and the back gate, we estimate $n_s/V_g$ to be $\sim 7\mathrm{x}10^{10}$ cm$^{-2}$/V. At the charge neutrality point $V_{CN}$, the device conductance attains its minimum. In an ideal graphene, this corresponds to nominally zero carrier density, where transport occurs entirely through the evanescent modes.[23] Away from $V_{CN}$, $g_m$ increases linearly with $V_g$, which corresponds to the "electron-doped" and "hole-doped" regimes.

The carrier mobility in the devices can be estimated from the slope of $g_m$-$V_g$ curve using the Drude definition $\mu = \sigma / n_s e$ (here $\sigma$ is the electric conductivity and $e$ is the electron charge). For our devices we obtained $\mu \sim 8,000 - 15,000$ cm$^2$/Vs, which attest to the material quality. Additional characterization was provided by the quantum Hall measurements at 8T magnetic field. Figure 1 (c) shows that the transverse resistance $R_{xy}$ exhibits plateau at $\dfrac{h}{(N+\frac{1}{2})\upsilon e^2}$, where h is the Planck's constant, $\upsilon$ accounts for spin and valley degeneracy ($\upsilon=4$ for graphene), and $N=0, \pm1, \pm2\ldots$ is an integer. In sharp contrast to the standard quantum Hall quantization at $\dfrac{h}{N\upsilon e^2}$, the anomalous "half-integer" plateau is a unique signature of the relativistic band structure of graphene. The latter clearly establishes our selection of SLG for the present study.

The micro-Raman spectroscopy was carried out using the Renishaw instrument. A Leica optical microscope with 50x objective was used to collect the backscattered light from the graphene samples. The Rayleigh light was rejected by the holographic notch filter with a 160 cm$^{-1}$ cut off frequency for 488 nm excitation. The graphene temperature was changed using the cold-hot cell with the step of 10±0.1 K. We varied the temperature of the sample in the range from 83K to 373K and recorded Raman spectra at 10K intervals.

Both *G* and *2D* peaks for SLG and BLG shift toward the lower frequency with the increasing temperature. The same trend was observed for the highly-oriented pyrolytic graphine (HOPG), which we used as a reference sample. One can introduce the *G* (or *2D*) mode temperature coefficient $\chi_{G,2D}$ through the expression for the peak frequency $\omega_{G,2D}=\omega^o{}_{G,2D}+\chi_{G,2D}T$, where $\omega^o{}_{G,2D}$ is the frequency of the *G (2D)* peak when the absolute temperature $T$ extrapolated to 0 K. The change of the Raman shift with temperature is a manifestation of the anharmonic potential constants, the phonon occupation number and the thermal expansion of the graphene 2D lattice.[22]

Figure 2 presents the temperature dependence of the $G$ peak position for BLG and HOPG. The inset shows the $G$ peak shape and position for SLG at two temperatures. The extracted $G$-mode temperature coefficient $\chi_G$ for SLG, BLG and HOPG are listed in Table I. The observed linear trend for the $G$ peak temperature dependence is consistent with the reports for other carbon-based materials. Ci *et al.*[24] observed the linear trend for $\chi$ of the radial breathing mode and the tangential stretching mode in the spectra from the double-wall carbon nanotubes in the range T=70K-650K. No deviation from the linear dependence was detected for the $D$ and $D*$ peaks in the spectra from CNTs or active carbon.[25] The non-linear term in the temperature coefficient for Raman peaks from diamond only appears at the high end of the temperature range T=293-1850K.[26]

In Figure 3 (a-b) we show the change in *2D* feature in the Raman spectrum from SLG and BLG when the temperature changes from 113K to 373K. Here we use the terminology proposed by Ferrari *et al.*[13] for the second-order band at ~2700 cm[-1]. The extracted values of the temperature coefficients $\chi_{G,2D}$ and zero-temperature frequencies $\omega^o_{G,2D}$ for $G$ and *2D* peaks in the spectra from SLG, BLG and HOPG are summarized in Table I. The absolute values of the *2D* peak thermal coefficients are larger that those of $G$ peak. The latter can be related to the fact that the *2D* feature is a second-order phonon peak. It is interesting to note that our result for $\chi_G$ of the reference HOPG sample exactly coincides with the value reported by Tan *et al.*[27] who also found $\chi_G$ = -0.011 cm[-1]/K. It is rather intriguing that $\chi_G$ of SLG is larger than that of BLG and HOPG.

The important conclusion for the nanometrology application of Raman microscopy is that the shift in the $G$ peak position due to the temperature change is comparable to the peak shift with the number of graphene layers $n$. Gupta *et al.*[14] found that the position of the $G$-band upshifts linearly relative to that of graphite with the increasing *1/n*. The overall shift, as $n$ changes from 19 to a single layer, is ~5-6 cm[-1]. Our measurements demonstrate that the change in the $G$ peak position as the temperature varied by approximately 300 K is about ~4-5 cm[-1]. The temperature variation of few hundred degrees can occur when the device is cooled for the low temperature measurements or as a result of the increased excitation laser power or application of the realistic bias voltages to the graphene-based devices. The metrology on the basis of *2D* band may be more robust with respect to temperature variation since the information about the number of layers is derived mostly from the shape of the *2D* feature rather than its position.

*Acknowledgements*

This work has been supported, in part, by the DARPA – SRC funded FCRP Center on Functional Engineered Nano Architectonics (FENA) and by the DARPA – DMEA funded UCR – UCLA – UCSB Center for Nanoscience Innovations for Defense (CNID).

**Table I:** Temperature Coefficients of the Raman Peaks for the Single-layer and Bi-layer Graphene

| material | peak | $\chi$ (cm$^{-1}$/K) | peak at 0K (cm$^{-1}$) | Temperature range (K) |
|---|---|---|---|---|
| single-layer graphene | G | -0.016 | 1584 | 83-373 |
| bi-layer graphene | G | -0.015 | 1582 | 113-373 |
| highly ordered graphite | G | -0.011 | 1584 | 83-373 |
| single-layer graphene | 2D | -0.034 | 2687 | 83-373 |
| bi-layer graphene | 2D | -0.066 | 2687 | 113-373 |

**Figure Captions**

**Figure 1:** (a) atomic force microscopy image of the graphene layer used for the device fabrication; (b) electrical conductance of the graphene device as a function of the applied gate bias; (c) transverse resistance of the graphene device as a function of the gate bias. Note that the value of the plateau confirms the relativistic band structure and selection of SLG.

**Figure 2:** Temperature dependence of the *G* peak position for BLG and HOPG. The inset shows the shape of the *G* peak and its shift for SLG.

**Figure 3:** Raman spectrum showing *2D* peak frequency at 113K and 373K for (a) SLG and (b) BLG.

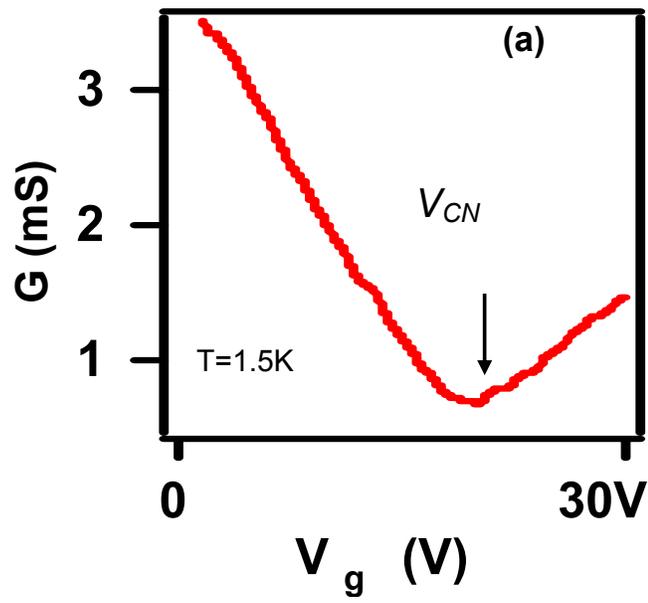

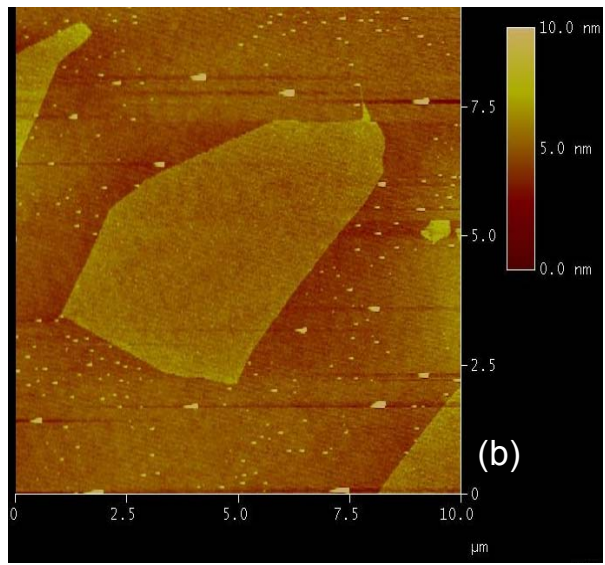

Figure 1 (a-b): Calizo et al., 2007

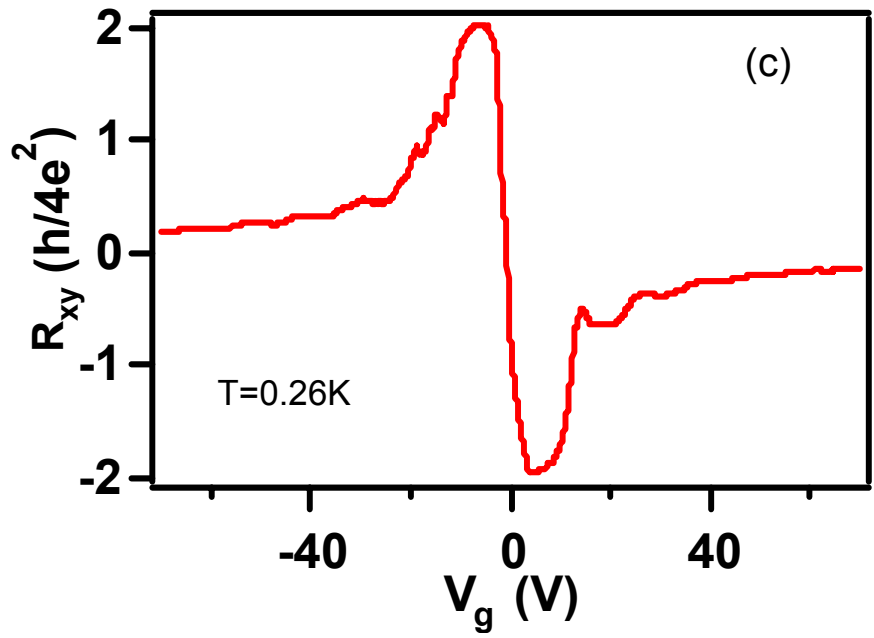

Figure 1 (c): Calizo et al., 2007

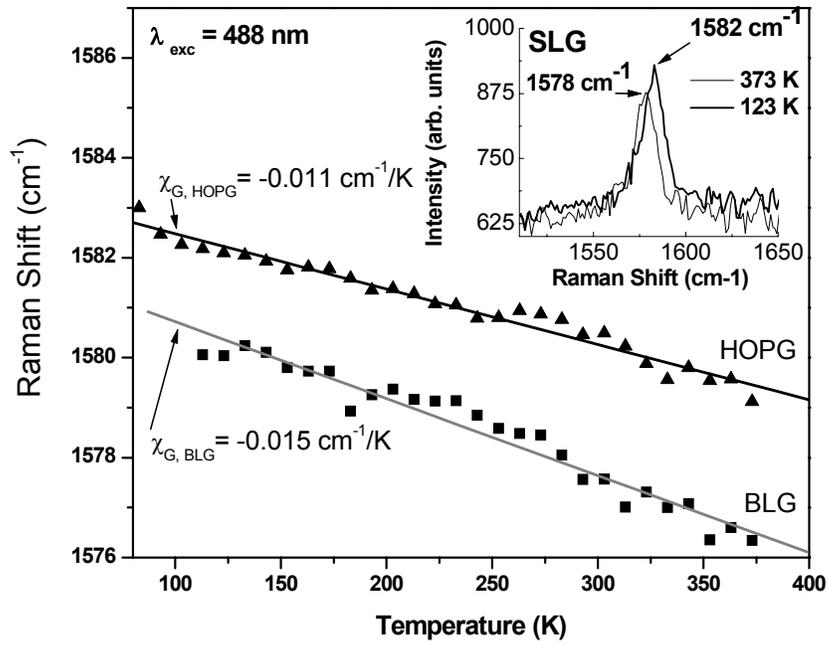

Figure 2: Calizo et al., 2007

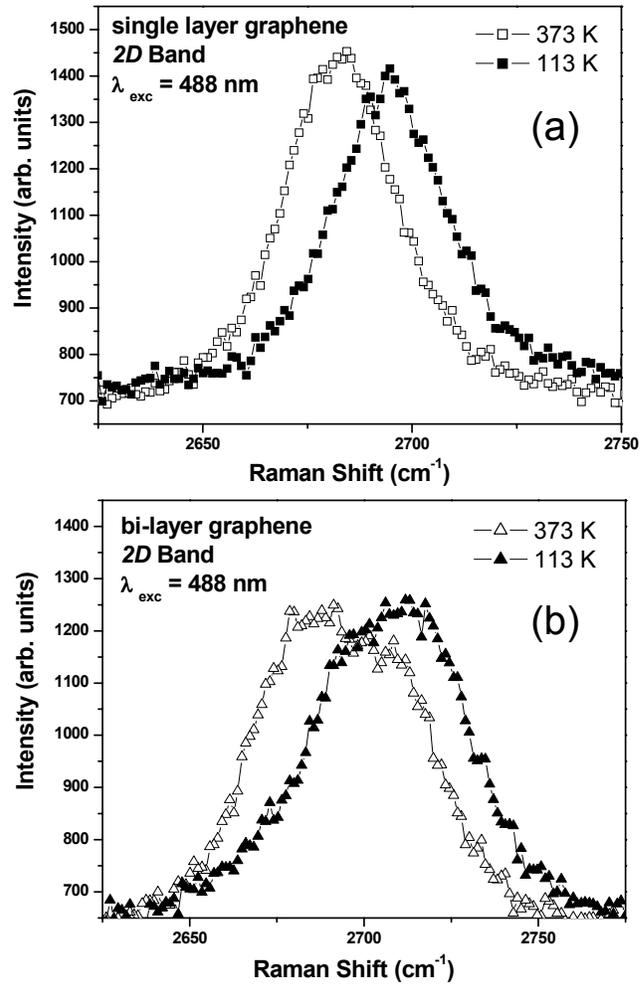

Figure 3: Calizo et al., 2007